\documentclass{jaa}
\usepackage{natbib}
\usepackage{xcolor}
\usepackage{multicol}

\usepackage{graphicx}
\usepackage{mathtools}

\usepackage{hyperref}
\hypersetup{
    colorlinks=true,
    linkcolor=blue,
    filecolor=magenta,      
    urlcolor=cyan,
    pdftitle={Overleaf Example},
    pdfpagemode=FullScreen,
    citecolor=red,
    }
\usepackage{graphicx}
\usepackage{url}

\begin{document}\sloppy

\title{Investigation of a Machine learning methodology for the SKA pulsar search pipeline}


\author{Shashank Sanjay Bhat\textsuperscript{1*}, Thiagaraj Prabu\textsuperscript{2}, Ben Stappers\textsuperscript{3}, Atul Ghalame\textsuperscript{3}, Snehanshu Saha\textsuperscript{4}, T.S.B Sudarshan\textsuperscript{5}, Zafiirah Hosenie\textsuperscript{3}
}
\affilOne{\textsuperscript{1}During the work affiliated to Raman Research Institute, Bangalore, India \\ \textsuperscript{*}Now with IBM India Private Ltd, Bangalore, India  \\
\textsuperscript{2}Raman Research Institute, Bangalore, India  \\
\textsuperscript{3}University of Manchester, Manchester  \\
\textsuperscript{4}APPCAIR, BITS Goa, Goa - 403726, India\\
\textsuperscript{5} PES University, Bangalore, India}


\twocolumn[{

\maketitle

\corres{ssbhat98@gmail.com}


\vspace{2em}
\begin{abstract}
The SKA pulsar search pipeline will be used for real time detection of pulsars. Modern radio telescopes such as SKA will be generating petabytes of data in their full scale of operation. Hence experience-based and data-driven algorithms are being investigated for applications such as candidate detection. Here we describe our findings from testing a state of the art object detection algorithm called Mask R-CNN to detect candidate signatures in the SKA pulsar search pipeline. We have trained the Mask R-CNN model to detect candidate images. A custom semi-auto annotation tool was developed and investigated to rapidly mark the regions of interest in large datasets. We have used a simulation dataset to train and build the candidate detection algorithm. A more detailed analysis is planned. The paper presents details of this initial investigation highlighting the future prospects.

\end{abstract}

}]


\doinum{12.3456/s78910-011-012-3}
\artcitid{\#\#\#\#}
\volnum{000}
\year{2022}
\pgrange{1--}
\setcounter{page}{1}
\lp{1}

\section{Introduction}
Modern Radio Telescopes, such as the Square Kilometre Array (SKA), generate data in the order of petabytes in their full scale of operation. Data volumes arise due to the extensive array size (197 antennas in South Africa and 1,31,072 antennas in Western Australia), multiple beams (up to about 1500), broadband digitisation (about 800 Msps), and from the demanding science requirements \citep{SKA_1,SKA_2}. Correspondingly there is a considerable high-speed signal processing: FFT, filtering, cross-correlations, beamforming, etc., employed using dedicated hardware components. Subsequent processing uses a mix of application-specific and non-deterministic algorithms, where statistical and experience-based decision making is indispensable.


Manual or semi-automated approaches will not be feasible when handling such volumes of data. We will have to employ automated and heuristic algorithms that can ease the classification of data, detection of patterns/objects in the data streams in real-time. Machine Learning, which is an actively evolving field, brings in a host of valuable practical techniques to efficiently handle the radio-telescope large volume data processing.

This paper investigates the use of a state of the art object detection algorithm called Mask R-CNN in order to detect accelerated binary pulsar's signatures in a pulsar search pipeline data stream. In a related work by \citep{Lyon_2016,Keith_2010,Bates_2012} candidate detection was performed on pulsar profiles. We demonstrate a novel approach of detecting candidate signatures in the middle of the processing pipeline. Additionally our effort for detecting binary pulsar signatures involving machine learning based approaches is novel. By determining the location, co-ordinates of the candidate we will be able to extract useful information for the sifting needed to sort the candidates. 


This paper is outlined as follows: Section \ref{sec:background} gives a brief background on machine learning. Section \ref{sec:search} introduces the SKA pulsar search pipeline \citep{Levin_2017}, the Fourier Domain Acceleration Search (FDAS)\citep{https://doi.org/10.48550/arxiv.astro-ph/0112006}  and the machine learning pipeline which was investigated. Section \ref{sec:future} provides a discussion about our main results. Section \ref{sec:futurepropects} talks about our future propects and Section \ref{sec:conclusion} concludes the work.

\section{Background on Machine Learning} \label{sec:background}
Machine Learning is the study of algorithms which can improve over time with experience and data. Machine Learning is classified into three categories : 1. Supervised Learning, 2. Unsupervised Learning and 3. Reinforcement Learning. 

The term Supervised Learning implies that the data is labeled. An Artificial Intelligence (AI) model creates a mapping between the features and output and uses this mapping function to predict values for new datasets. Support Vector Machines \citep{708428}, Decision Trees \citep{Decision}, K-Nearest Neighbors \citep{Cunningham_2022} are some of the widely used supervised machine learning algorithms. 

Unsupervised machine learning signifies AI models which creates a mapping function by clustering and grouping similar data. K-means clustering \citep{10.1007/978-3-540-87479-9_3}, Expectation Maximization clustering \citep{543975}, Auto encoders 
 \citep{https://doi.org/10.48550/arxiv.2003.05991} are some of the widely used unsupervised machine learning algorithms. 

Reinforcement Learning works on the concept of AI agent which observes its surroundings and performs an action. Based on the action taken it is either rewarded or penalized \citep{Reinforcement}. Over time the AI model will aim towards maximizing the reward function. 

\begin{figure}
  \centering
  \includegraphics[scale=0.25
 ]{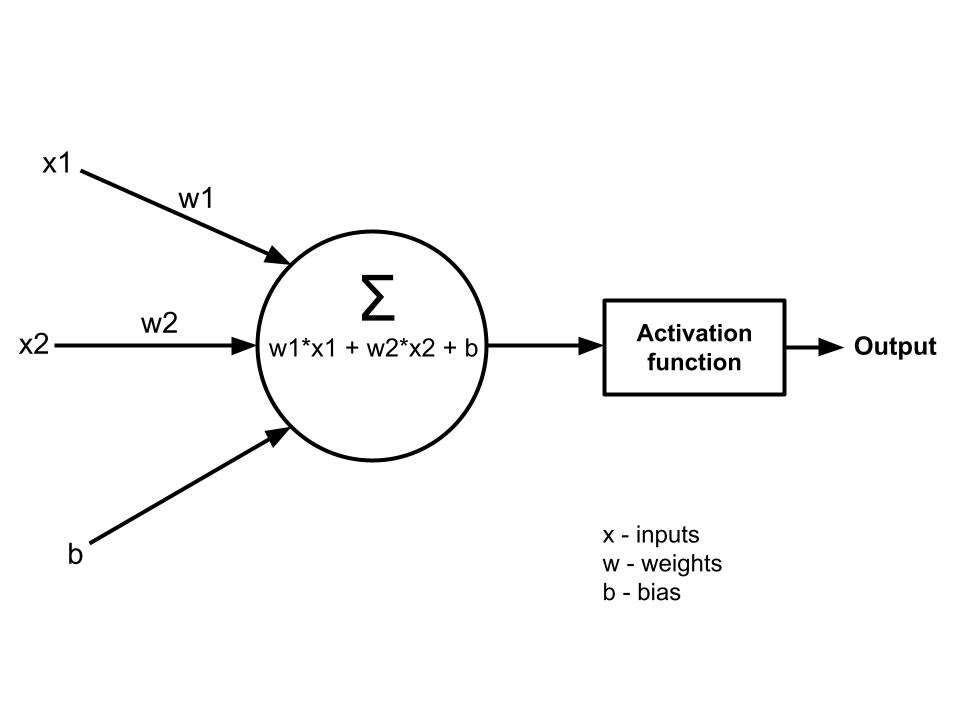}
  \caption{The structure of an artificial neuron having two inputs, two weights, a bias and an activation function}
  \label{fig:neuron}
\end{figure}

In most of the above three categories of machine learning the underlying workhorse is a neural network. A neural network is primarily composed of neurons \citep{Neurons}. A neuron takes a sum of weighted inputs along with a bias and subjects this output to an activation function. The activation function determines if a neuron has to switch on or off. Figure \ref{fig:neuron} shows a general structure of a neuron with two inputs and a bias. Neural Network based approaches in some cases create the mapping between the input data and its corresponding labels.


\begin{figure}
  \centering
  \includegraphics[scale=0.30
 ]{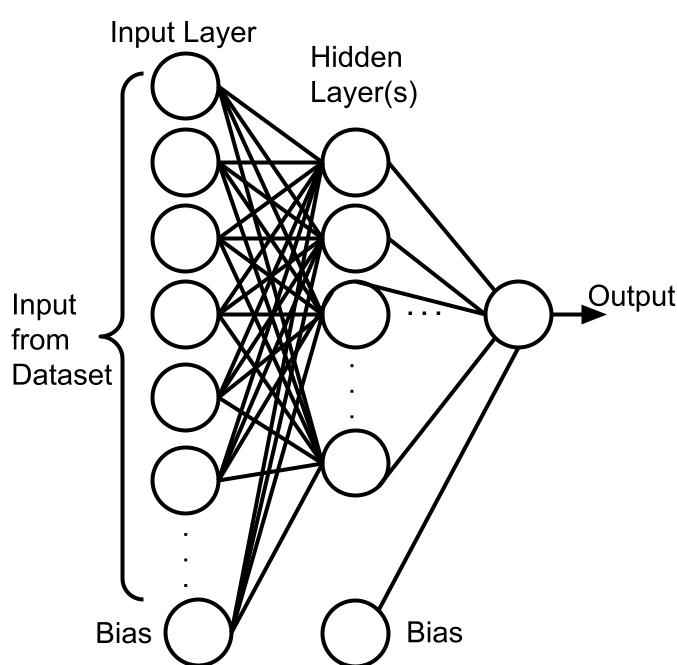}
  \caption{An Artificial Neural network having multiple neurons, one input layer, one hidden layer and one output. For complex problems the neural network can have multiple hidden layers. For multi class classification problems there are multiple neurons in the output layer.}
  \label{fig:ann}
\end{figure}

Neural networks can be effectively used for modeling the behaviour of a function. But in case of complex functions more number of neurons in the form of layers will have to be added. A neural network with multiple hidden layers is called a deep neural network. Figure \ref{fig:ann} shows an artificial neural network with one hidden layer.

In case of pulsar searches we will have to deal with large amounts of image based data and the processing will have to be done in real time \citep{https://doi.org/10.48550/arxiv.1810.06012}.  Lyon et al, 2018 in their paper introduces a real time pulsar candidate detection pipeline. 


When dealing with image data, we use a special case of an Artificial Neural Network called the Convolutional Neural Network (CNN) \citep{NIPS2012_4824}. The feature extraction in a CNN happens with a repeated set of convolutions and pooling. Convolutions are performed via filters to produce feature-maps. The pooling layer is responsible for reducing the spatial size of the feature maps. The feature map, additionally helps in reducing subsequent computation required.


\section{SKA Pulsar Search pipeline} \label{sec:search}
\begin{figure}
  \centering
  \includegraphics[scale=0.5]{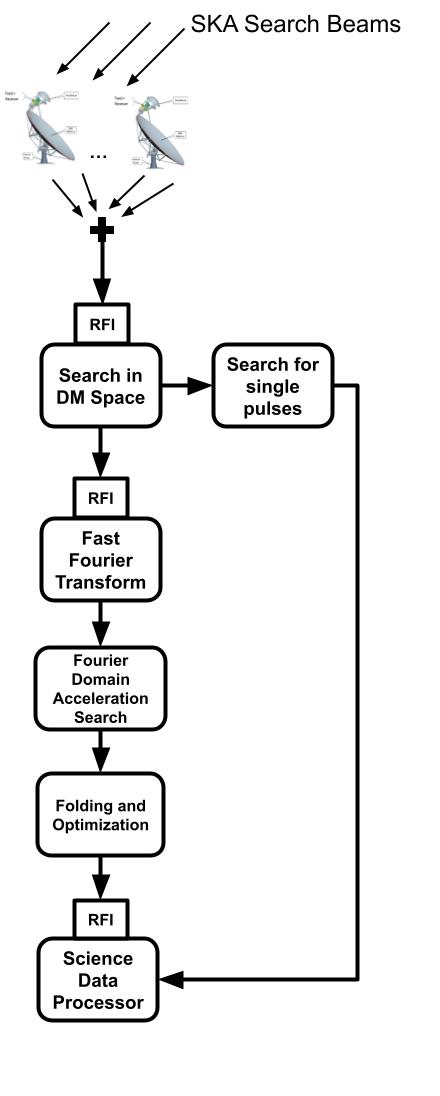}
  \caption{Schematic of the pulsar search processing work flow in SKA. FPGA based acceleration architecture investigated for 1) RFI mitigation, 2) dedispersion, 3) FFT, 4) acceleration search and 5) candidate folding optimisation modules.}
  \label{fig:pulsar_pipeline}
\end{figure}

One of the complex tasks of the SKA pulsar engine shown in Fig \ref{fig:pulsar_pipeline} is to search for pulsars in binary systems where the apparent frequency of the pulsar is changed significantly during the observation.

\begin{figure}[htbp]
  \centering
  \includegraphics[width=0.5\textwidth]{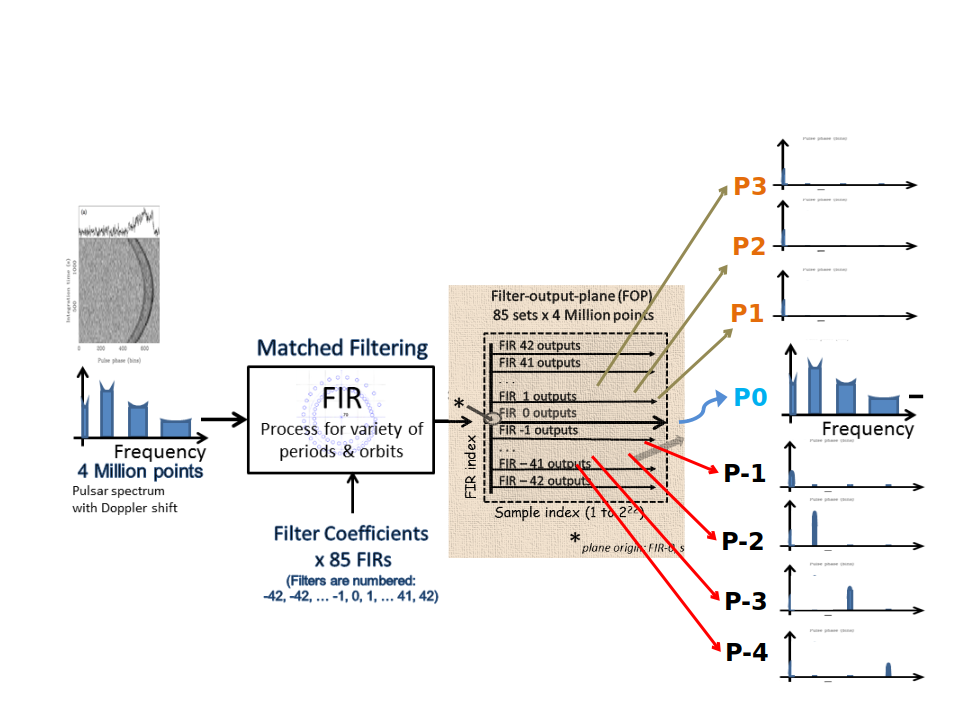}
  \caption{Main processing in FDAS consists of matched filtering to deconvolve the fundamentals and their harmonics \citep{fdas}}
  \label{fig:fdas_day1}
\end{figure}

A Fourier domain acceleration search (FDAS) algorithm \citep{Ransom_2002} is being tested for the SKA application. The frequency changes manifest in the Fourier space as \emph{sinc} functions convolved with an FIR response. In order to deconvolve these signals, a set of optimised matched filters are constructed and used in the FDAS module (Fig \ref{fig:fdas_day1}). 

\begin{figure}[htbp]
  \centering
  \includegraphics[width=0.5\textwidth]{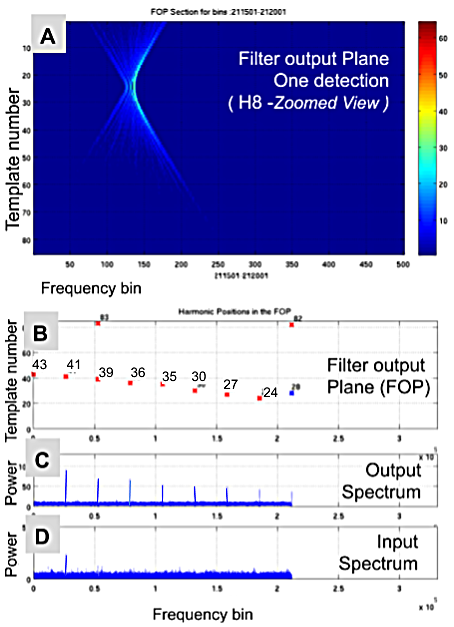}
  \caption{A: An intensity image of the  filter output plane around one of the harmonic positions. B: Signal's harmonic positions and a few spurious detections. C: Spectrum after the matched filtering. D: Input spectrum. In this illustration we have used the double pulsar test vector \citep{fdas}.}
  \label{fig:fopview}
\end{figure}

The FDAS module receives the RFI mitigated complex spectra generated from the dedispersed time series. Each spectrum is passed through a set of 85 matched filters, each analysing a unique acceleration-period range in the search. Outputs from the filters are detected to obtain the power series and saved in a 2-dimensional data array known as a filter-output-plane (FOP) (Fig \ref{fig:fopview}).


For SNR optimization across the search parameter space, 84 filters are required to search across pulsar periods up to 500 Hz with +/-350 $ms^{-2}$ acceleration range. A typical data size of 4 M samples, which corresponds to an observation time of about 10 minutes \citep{fdas}.


Our focus in the work is on the Fourier Domain Acceleration Search module. We have investigated the use of Mask R-CNN to detect candidates of accelerated binary pulsar signatures.


\subsection{Fourier Domain Acceleration Search}
The intermediate output product of the Fourier Domain Acceleration Search (FDAS) processing when viewed as a particular image intensity mode a distinct shape (Hourglass, Butterfly) is observed. These  butterfly shapes comes from the particular arrangement of the filters, and when an accelerated  binary pulsar fundamental or harmonics signals gets deconvolved through them. The intensity, location and inclination of these shape / patterns give us information about these candidates. 




Conventional processing in this stage, involves sifting through the FOP array and performing a harmonic summing And thresholding. Additionally, basic CNNs (Convolutional Neural Network) perform classification,  but we investigate an improved version of CNN, known as the Mask-RCNN that can identify the patterns and provide a mark the regions (bounding box) where pattern features are identified, Typically the CNN is computationally expensive to implement. In our case due to the use of accelerators additional computation steps for implementing convolutions is less significant.


The R-CNN and Fast R-CNN algorithm \citep{https://doi.org/10.48550/arxiv.1504.08083} which uses a selective search algorithm could not be  used for real life deployments as there was a performance bottleneck. The Faster R-CNN algorithm \citep{https://doi.org/10.48550/arxiv.1506.01497} could not be used as we needed a mask to be drawn for every detected object. Due to these reasons we chose Mask R-CNN to perform candidate signature detection. Our investigation is to see if the Mask R-CNN suite can be used to train and recognize these image patterns. We have identified the public domain availability of Mask R-CNN from matterport github repository.(\hyperlink{https://github.com/matterport/Mask_RCNN}{Repository Link})\footnote{\url{https://github.com/matterport/Mask_RCNN}} 


\subsection{Mask R-CNN}
\label{sec:mask}
Mask R-CNN \citep{https://doi.org/10.48550/arxiv.1703.06870} is a small, flexible generic object instance segmentation framework. It not only detects targets in the image, but also gives a high-quality segmentation result for each target. It is extended on the basis of Faster R-CNN, and adds a new branch for predicting an object mask which is parallel with bounding box recognition branch. 
For object classification we have used the default ResNet101 \citep{https://doi.org/10.48550/arxiv.1512.03385} as our classfication network. For the Detection network, we have used a Region Proposal Network which was proposed with the Faster R-CNN algorithm. Feature Extraction in the Mask R-CNN happens via the Feature Pyramid Network (FPN) algorithm. The FPN algorithm was designed to handle images of various sizes and shapes by keeping speed, memory and accuracy in mind. The algorithm consists of both a top down and bottom up approach for producing feature maps. The bottom up network features a regular convolutional neural network for extracting features from images. Every layer in the bottom up pyramid reduces spatially allowing for high level structures to be determined, thereby increasing the semantic value. The top down network provides a pathway to construct high resolution layers from the semantic rich layer. Additionally there are lateral connections provided to help the detector better detect the object locations. This algorithm has shown significant improvements over the other state-of-the-art approaches \citep{https://doi.org/10.48550/arxiv.1612.03144}. Mask R-CNN defines the loss function \footnote{\url{https://en.wikipedia.org/wiki/Loss_function}} as follows:
\begin{equation}
    L = L_{cls} + L_{box} + L_{mask}
\end{equation}
Where $L_{cls}$ denotes the classification loss, $L_{box}$ denotes the bounding box loss and $L_{mask}$ is the average binary cross entropy loss, only including k-th mask if the region is associated with the ground truth class k.

\subsection{Basic Tests with Mask R-CNN}
Mask R-CNN is available as a github repository from matterport.
The github repository is cloned onto our local system following the setup procedure given in the github repository 

\begin{figure*}[!ht]
  \centering
  \includegraphics[height= 8cm,width= \textwidth]{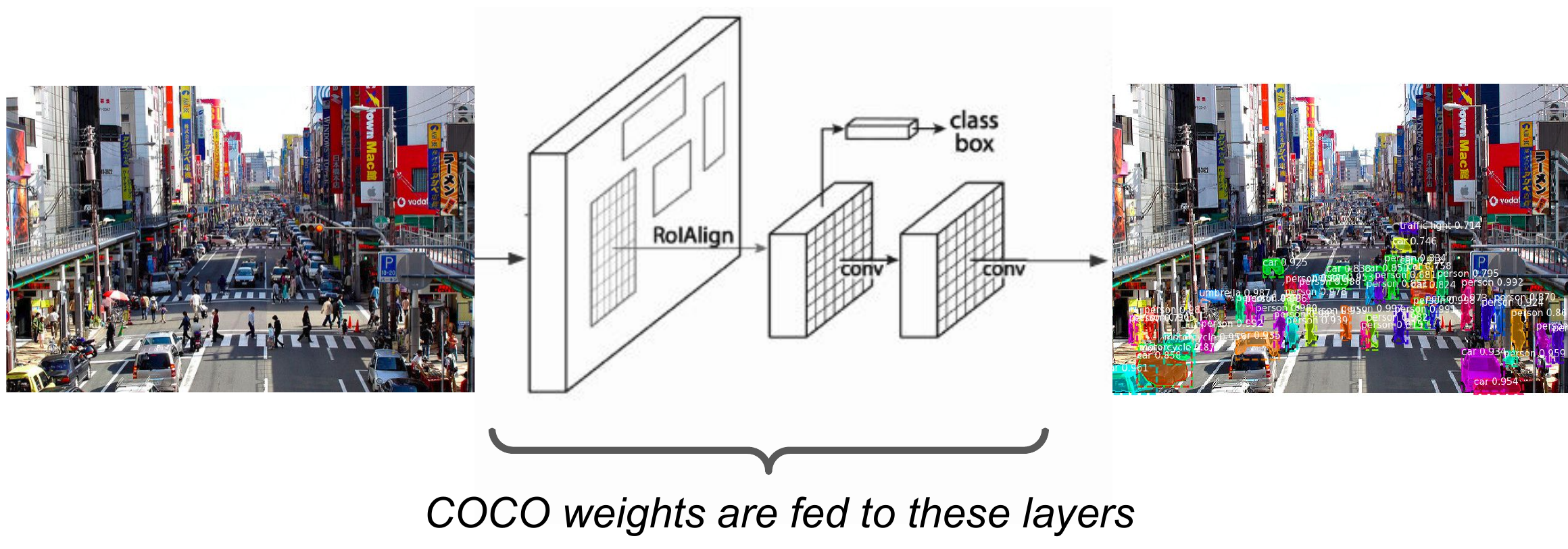}
  \caption{As a basic step we took the default MS-COCO weights along with Mask R-CNN. MS-COCO weights are provided along with the Mask R-CNN repository. A city image was taken and subjected to this algorithm. We observed that to one extent of the resolution of the image it could detect objects, draw bounding boxes on each of the detected along with a confidence score and additionally draw a mask for each of the detected objects. }
  \label{fig:mrcnn-city}
\end{figure*}

At the time of this work, the default Mask R-CNN available from the repository has been trained on 80 different objects from the MS COCO \citep{https://doi.org/10.48550/arxiv.1405.0312} dataset. Figure \ref{fig:mrcnn-city} is an example illustration to show the default detection capability of Mask R-CNN. Image shown is a city street and the detections are highlighted with bounding boxes. 

\begin{figure}
  \centering
  \includegraphics[width=0.5\textwidth]{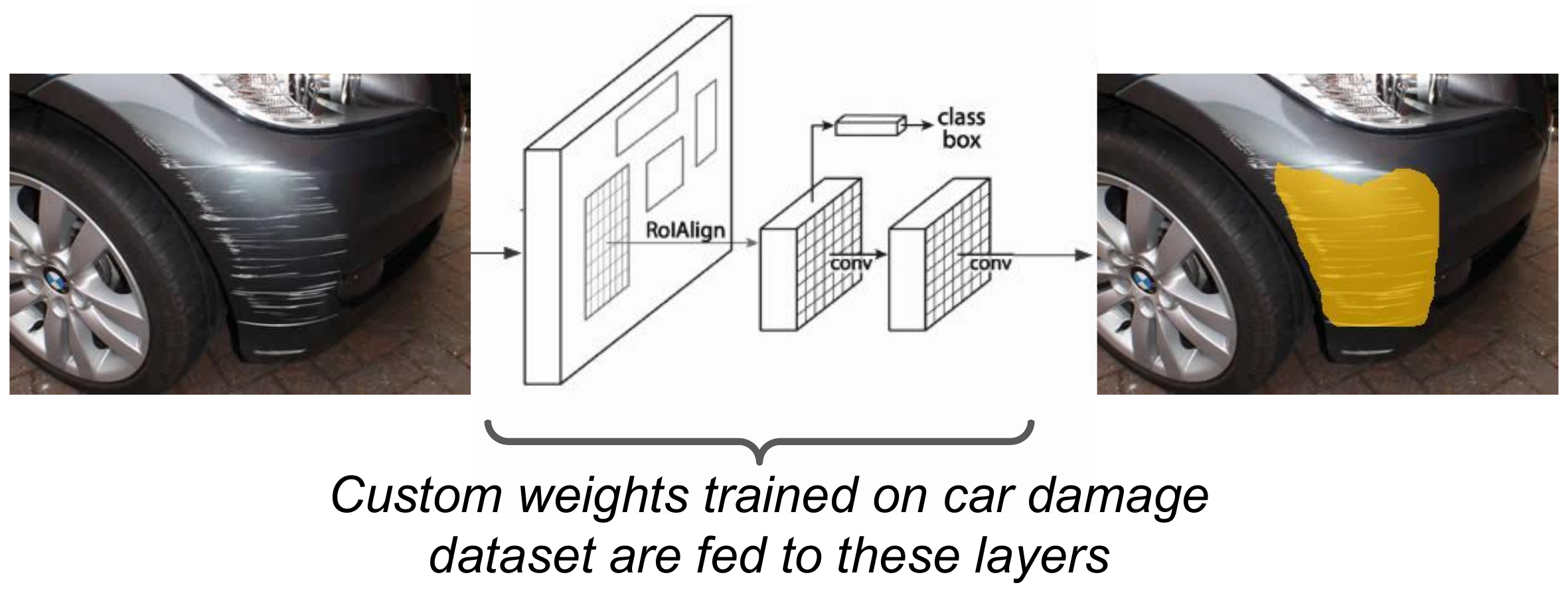}
  \caption{Before getting into candidate detection, we trained Mask R-CNN on a custom car damage dataset. Input images along with their respective masks were taken as input to the model. The custom weights were extracted and were tested on test images. This figure shows the result on one of the test images which was taken.}
  \label{fig:mrcnn-car}
\end{figure}

In addition, in order to understand the training procedure of Mask R-CNN we have experimented it to detect damages in a set of car images. For this purpose we collected sample images from an external github repository (\hyperlink{https://github.com/priya-dwivedi/Deep-Learning/tree/master/mask_rcnn_damage_detection/customImages}{Repository Link})\footnote{\url{https://github.com/priya-dwivedi/Deep-Learning/tree/master/mask_rcnn_damage_detection/customImages}}. It was required to annotate the images using Visual Geometry Group (VGG) annotator. We then trained the Mask R-CNN suite using the standard procedures described in github. The loss function for this experiment remains the same as the standard loss function described in section \ref{sec:mask}. For this experiment we havent performed any predictive success analysis \footnote{\url{https://en.wikipedia.org/wiki/Predictive_analytics}}. Figure \ref{fig:mrcnn-car} shows the result from training Mask R-CNN on the car damage dataset. 

The training of the neural network is an important task and this procedure is described in the next section.

\subsection{Training a neural network}
Every neural network requires a good dataset (both qualitatively and quantitatively). A dataset is split into three parts i.e., a training set which is used in the neural network, a validation set which is used for preventing overfitting and a testing set to evaluate the performance of a neural network. During the training of the neural network there is a specific loss function which is defined. A loss function can either be predetermined or a custom user defined. The loss function gives us an approximate idea about how well an algorithm models a given data. A neural network updates its weights with the backpropagation algorithm. A forward pass in a neural network is defined as one run starting from the input layer until the output layer. A combination of one forward propagation and one backward propagation is called an epoch. During the training cycle the algorithm is subjected to the training data and is validated with the validation dataset. When the validation loss begins exceeding the training loss we can safely conclude the training process and extract the weights. The weights are a crucial database produced at the end of the training process which will help in making appropriate connection between the neurons. The weights will be provided to the inferencing logic. 


\subsection{Adapting Mask R-CNN for candidate detection}
As discussed before,  The default Mask R-CNN is used for detecting generic objects. We will have to train the Mask R-CNN to produce a new set of weights for detecting the candidate signatures (hourglass/butterfly patterns). This process involves four steps :
\begin{itemize}
    \item Preparing the dataset
    \item Image Annotation
    \item Training
    \item Testing
\end{itemize}

\begin{figure}
  \centering
  \includegraphics[width=0.5\textwidth]{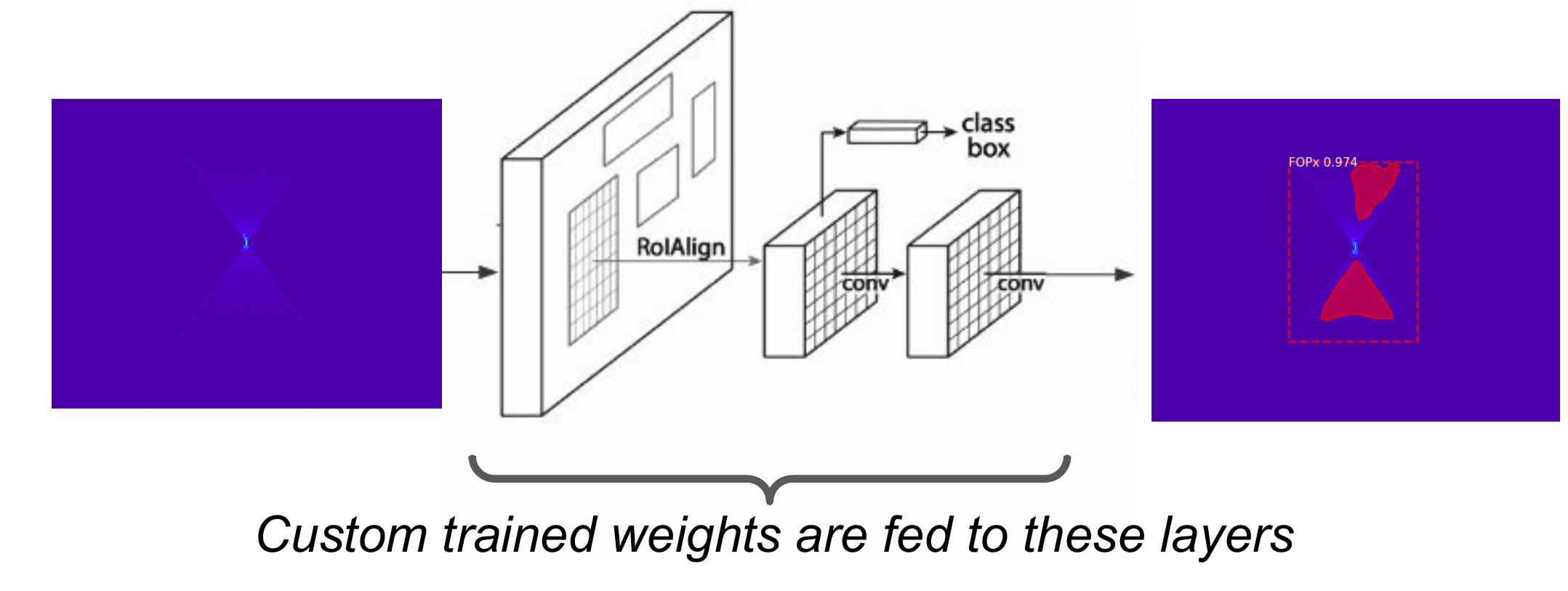}
  \caption{A Bird's eye perspective of how Mask R-CNN will work for our task. Figure shows the ideal input and output scenario for a candidate signature image. The custom dataset will be annotated and then trained on Mask R-CNN in order to get a custom weights file. This weights file will be used for detecting signatures on a new set of test images.}
  \label{fig:mrcnn}
\end{figure}

\subsection{Dataset Preparation}
In order to train the neural network, we have used different sets of binary pulsar signal data. We had a choice of using real observational data from a telescope or using simulated data. A telescope data will have the presence of interfering signals which is undesirable for initial training of the machine learning (ML) suite. In order to maintain a controlled test/training environment we have considered to use simulated datasets. Such simulated data are produced using a mock pulsar signal generation tool called  \hyperlink{https://sigproc.sourceforge.net/sigproc.pdf}{SIGPROC}\footnote{\url{https://sigproc.sourceforge.net/sigproc.pdf}} fake utility. Fake is a command line program written to create test data sets containing pulses hidden in Gaussian noise background. Various (38 different) datasets for the training purpose were produced by modifying the -period, -width, -snrpeak, -binary, -bper parameters of the fake utility. We have used data files produced with accelerations (+/- $500 ms^{-2}$, +/- $250 ms^{-2}$, +/- $25 ms^{-2}$ and 0 $ms^{-2}$ ), periods (2 s and 0.002 s), pulse duty cycle ratios (0.4, 0.2, 0.1 and 0.05), a constant SNR value of 80 and dispersion measure of 1.0. The simulated data obtained are in a filterbank format. The sampling period was set to 64 microseconds and the observation period was set to 536 seconds (close to real observation parameters). The data files are first dedispersed and channel collapsed to form a time series of 8 million samples.\footnote{\url{https://www.jb.man.ac.uk/research/pulsar/Education/Tutorial/tut/tut.html}}

\begin{figure}
  \centering
  \includegraphics[width=0.5\textwidth]{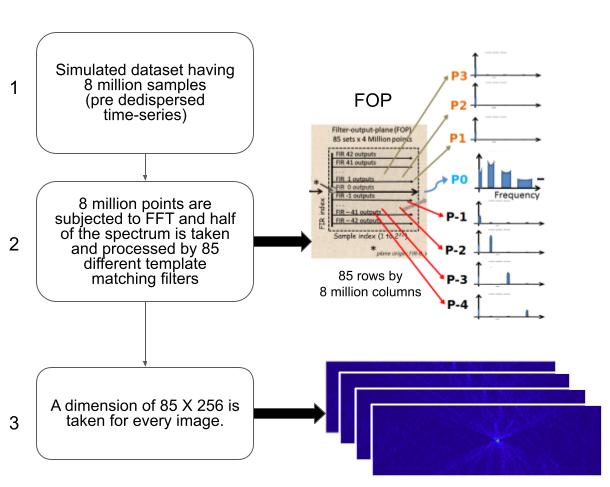}
  \caption{The taskflow involved in the image dataset preparation. GNU Octave scripts were developed and used for this purpose.}
  \label{fig:preparation}
\end{figure}

\begin{figure}
  \centering
  \includegraphics[width=0.5\textwidth]{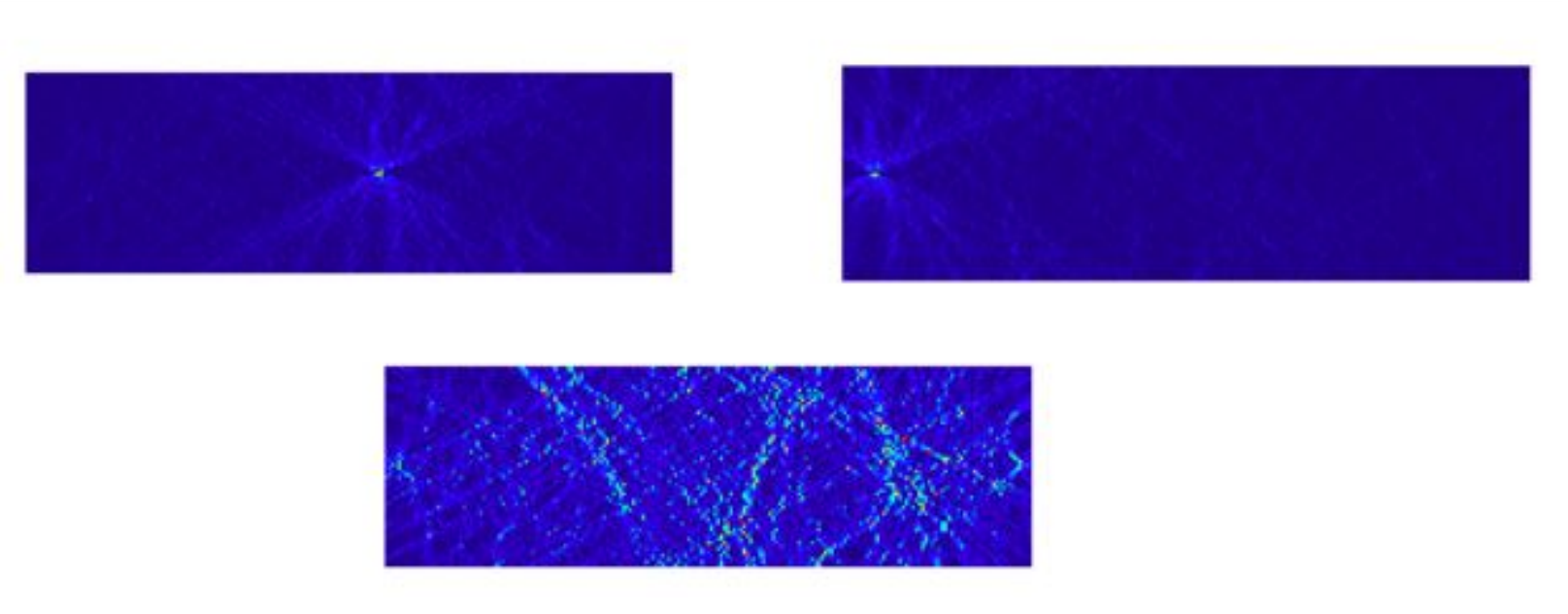}
  \caption{The first image on the top left shows a candidate image with a high intensity. We have taken the size of the image to be 85 * 256 pixels. The candidate signatures in the shape of a butterfly are obtained after subjecting the candidate array to 85 different acceleration filters. The second image in the top right shows the same candidate but it is shifted to the corner. The third image in the bottom shows the candidate image dominated with noise.}
  \label{fig:dataset}
\end{figure}

The time series data is subjected to 8 million point Fast Fourier Transform (real to complex). The resulting 4 million points after the transform are convolved with 85 templates to get a filter output plane (FOP), which is an array of dimension 85 rows by 4 million columns. A schematic of the data preparation flow is shown in Fig \ref{fig:preparation}.

The FOP array is subjected to a sliding window approach to extract images used for training the Mask R-CNN model. The images are of size 85 rows by 256 columns (similar to an image having 85x256 pixel size). Since the FOP is a very large array, there are many regions in the array without any significantly useful data for the training purpose. So we have selected regions where there are more likelihood presence of signals and noise combinations (starting from fundamental frequency locations to their multiple harmonic positions). This sliding window approach helped us to get 886 image segments (85x256 arrays) from the original FOP. The number 886 is an arbitrary choice and adequate to select a smaller training dataset. From this dataset we have picked up 50 good quality images for the training purposes and 34 random set of images for testing. Figure \ref{fig:dataset} shows the images obtained by processing different datasets that are having varying signal strengths (Signal to Noise Ratio). After the images are obtained they are subjected to image boundary annotation, as discussed in the next section.

\subsection{Image Annotation}
The Mask R-CNN model requires annotating the images using an image boundary annotation tool. The developers recommend using a standard VGG annotator. Initially we have used this standard tool for annotating the images however we found this tool required more human involvement to edit the boundaries and appeared cumbersome to annotate large number of images . Hence we explored developing our own custom annotator tool with features to annotate the images a little faster than the VGG tool. The annotation speed performance was achieved by automatically drawing the six sides (edges) needed for bounding the butterfly image boundaries. As mentioned earlier,for VGG it was required to draw each of the six sides manually which was consuming more time. The custom annotator was made use of to test short bounding box based annotations. 
The working principles of the annotation tools are presented below for a comparison.

\begin{figure}
  \centering
  \includegraphics[width=0.5\textwidth]{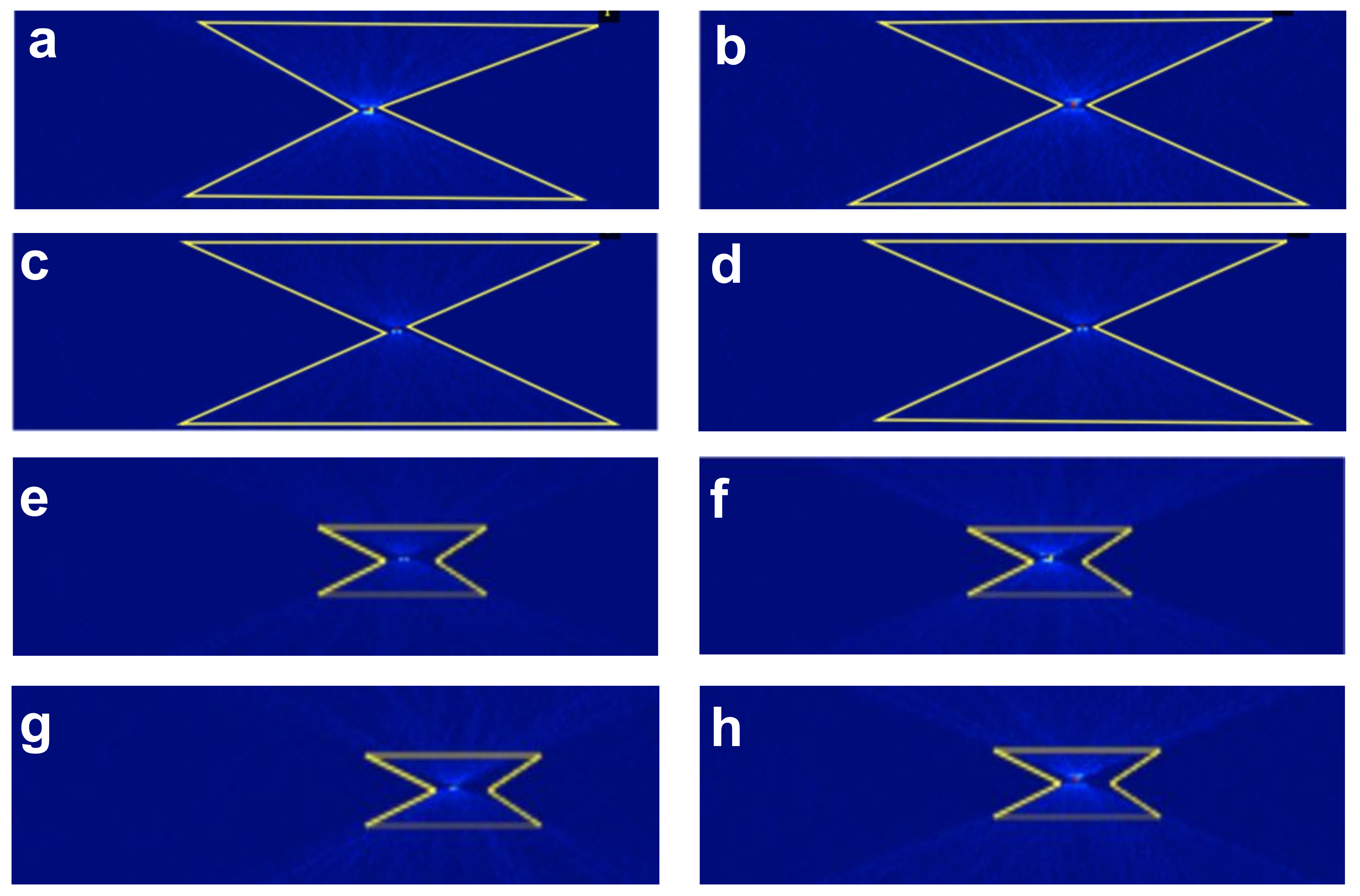}
  \caption{Example images from the training dataset. Subfigures a to d were achieved using the web based VGG annotator tool. Subfigures e to h were made using our custom annotator developed from this work. This effort helped us to analyze the effectiveness of the extent of the annotation. The yellow lines in all the subfigures mark the region of interest. }
  \label{fig:vgg}
\end{figure}

VGG Annotator developed by the visual geometry group\footnote{\url{https://www.robots.ox.ac.uk/~vgg/software/via/}} is very useful to draw multiple segment annotation boxes which are required for complex images such as humans, animals, buildings, cars etc \citep{dutta2019vgg}. After the manual annotation, the co-ordinates of the bounding box are saved in a JSON \footnote{\url{https://en.wikipedia.org/wiki/JSON}} file format. We have produced one set of (50 images) training image dataset using the VGG annotator. We annotated our training data set (consisting of 50 images) manually using a six point reference for the mask and saved the co-ordinates in a json format (Fig.\ref{fig:vgg}a-d). 

The custom annotator was developed to speed up the annotation process\footnote{\url{https://drive.google.com/drive/folders/1JOeRlDB4ckhhG9QeNID1qK0eMF19Wbo8}}. It is based on python and javascript. The code can be executed in a web browser. On loading the image the user needs to select the mid point of the butterfly image and the tool will automatically draw the six edges of the bounding boxes. We have pre-estimated and fixed upper limits for the dimension of the bounding box and thereby the code could automatically complete the annotation process. Upon selecting (with a mouse-click) the butterfly image's center, the annotator calculates the six nodal points of the bounding box with respect to the center point chosen by the user. We have used this tool to create a second set of training dataset which is essentially the same as the one produced by the VGG annotator but with shorter wings (Fig.\ref{fig:vgg}e-f).


After the annotation was completed we pass the dataset for training. The training process is explained in the next session.

\subsection{Training}
The Mask R-CNN suite comes with a toolflow for training. It is a computationally intensive task and we have installed the required programs in the Google Colab \footnote{\url{https://colab.research.google.com/}} cloud environment with GPU-based acceleration. While performing the training we have saved the weights-file produced after every epoch. The number of epochs required is not known apriori. So we have arbitrarily started the training process with the limit set to six epochs. But we found that the training was converging much before the third epoch. 

The loss curves is a progress indicator of the training process. Typically the flattening  of the loss curve around lower values indicates that the training is complete and it can be terminated. At each new epoch the same dataset is fed, but in a shuffled order to the network in order to build an effective mapping function for the weights. 

\begin{figure}
  \centering
  \includegraphics[scale=0.7]{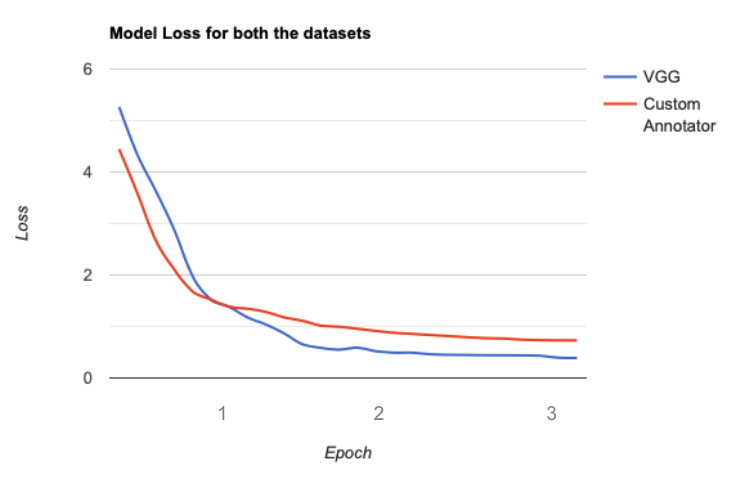}
  \caption{We subjected the Mask R-CNN model to train on 50 high resolution images. The images were annotated with two different extent of annotation. The loss curve obtained for two different kind of annotation scheme datasets (VGG and custom) are shown. VGG annotator was used to produce full extent annotations (Fig.\ref{fig:vgg}a-d) and the custom annotator was used for producing short annotations (Fig.\ref{fig:vgg}e-h). The loss curve flattened after the second epoch on both cases. Training was terminated after the third epoch. Marginally lower loss was observed are observed for the full extent training set.
  }
  \label{fig:loss}
\end{figure}

Figure \ref{fig:loss} shows the loss curves obtained for the training process. We can observe that after epoch two the loss curves are flattening. It can also be observed that two of our datasets (full extent annotated by VGG and short extent annotation by custom annotator) have shown similar trends. Since the loss curves started flattening at the second epoch we terminated our training process at the next (3rd) epoch. We utilized the transfer learning \footnote{\url{https://en.wikipedia.org/wiki/Transfer_learning}} method to train our Mask R-CNN suite between each of the epochs. The pretrained COCO weights \citep{https://doi.org/10.48550/arxiv.1405.0312} 
were used as the initial base weights file for the training. The model was trained with a learning rate of 0.001 on both datasets. The learning rate is a tuning parameter that determines the step size at each iteration while moving toward a minimum of a loss function. A larger learning rate signifies a higher chance of convergence at a local minima. A smaller learning rate signifies a higher chance of convergence at a global minima. The local and the global minima are referred to the loss function.

As mentioned before the training is computationally intensive and we have used Google Colab which is equipped with a Nvidia T4 GPU. Mask R-CNN has a specific configurable parameter which is known as \texttt{IMAGES\_PER\_GPU}. We found that by limiting the number of images to be processed by the GPU as two, the computation went faster. Figure \ref{fig:loss} shows the model loss curves produced during the training process of the two datasets.


\subsection{Results from our experiment}
\begin{figure*}[!ht]
  \centering
  \includegraphics[width=0.9\textwidth]{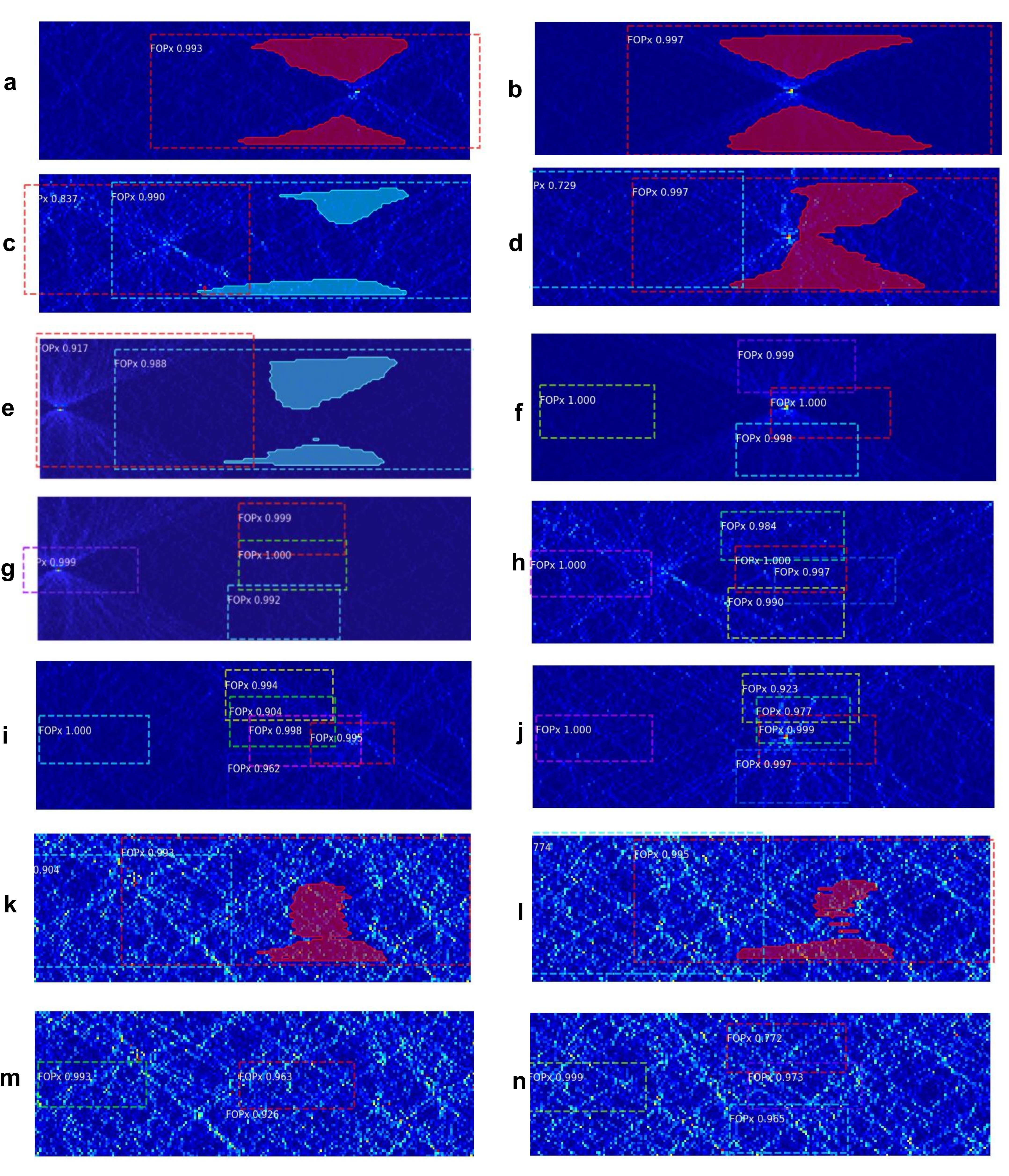}
  \caption{Ensemble of detections corresponding to images with low levels (a,b,..) of noise to the images with high levels (..,m,n) of noise are shown. The test-images had an acceleration value of 500 $ms^{-2}$, period of 2 s, pulse duty cycle ratios of 0.4, 0.05 and with a constant SNR of 40. At low noise levels the regions of interest were detected with high accuracy. At high noise levels there were multiple false detections.  Different colored bounding boxes (dashed lines) indicate different instances of the detected regions.}
  \label{fig:result2}
\end{figure*}
The training process generates the weights file for inferencing. The weight file is collected and passed onto the Mask R-CNN inferencing logic. We have considered 34 images for testing purposes. Out of the 34 images, four images had a high SNR, 13 images had a medium SNR and 17 images had very low SNR. Figure \ref{fig:result2} shows a subset of the images passed through the Mask R-CNN inferencing logic. As these were the first experimental results elaborate statistics were not collected. However we have the following qualitative analysis: 


\begin{itemize}
      \item The full length annotation based training was able to identify the regions and mark them with high confidence scores. The masks appeared only for very high SNR cases.(Fig \ref{fig:result2} a,b) 
      \item When the butterfly pattern is located around the edges of the image, we observed multiple detections having overlapped regions. (Fig \ref{fig:result2} e,g,i)
      \item  With low SNR cases false detections appeared at multiple places. (Fig \ref{fig:result2} k,l)
      \item Short length annotation based training resulted in higher false detections in the high SNR cases. 
      \item The medium SNR images also gave multiple detected objects with overlapped regions. (Fig \ref{fig:result2} h,i,j)
      \item  The low SNR cases also gave multiple false detections. (Fig \ref{fig:result2} m,n) 
      \item  In each of the above three cases the mask was not drawn over the detected objects. 
      \item Image pixel size of the training region seemed to play a major role in the detection and loss function estimation.
      
      

\end{itemize}

\section{Discussion} \label{sec:future}
The current design of SKA pulsar search is likely to use machine learning approaches in a variety of places, for example, for RFI detection and pulsar candidate identification \citep{Lyon_2016}.  We have investigated a machine learning approach for a new application in the search processing (Fourier Domain Acceleration Search (FDAS)) where conventional thresholding based approaches are usually followed. We have trained the network for a single feature detection. Usually the Mask R-CNN applications detect objects with well-defined boundaries where the object edges are well defined (sharp) with regard to the background. In our case the object has a fuzzy boundary but with a recognizable butterfly pattern for the human eyes. We have carried out this investigation to detect such fuzzy objects as a research work within the SKA pulsar search activity. Our work demonstrated the ability to train a network to detect such fuzzy objects. It has opened up further possible studies where we can do more quantitative studies on the algorithm and its computational performance improvements for the pulsar search work in general.
In addition we have also looked at the limitations of the standard annotation tools and also studied the extent of annotation required using a semi-automated custom annotation tool. There is scope for further enhancement of this custom annotation tool. The present training method used a single feature, by including a few extended features to the training (some noise patterns) and code enhancements, we will be able to determine the location, inclination and intensity of the candidates more precisely. Such information will enable simplifying usual complexities associated with subsequent candidate sifting (Sorting) and related processing. Thus our detection pipeline can be further improved to provide higher level of information during the search.


\section{Future Prospects} \label{sec:futurepropects}
\begin{figure}
  \centering
\[  \includegraphics[width=0.5\textwidth]{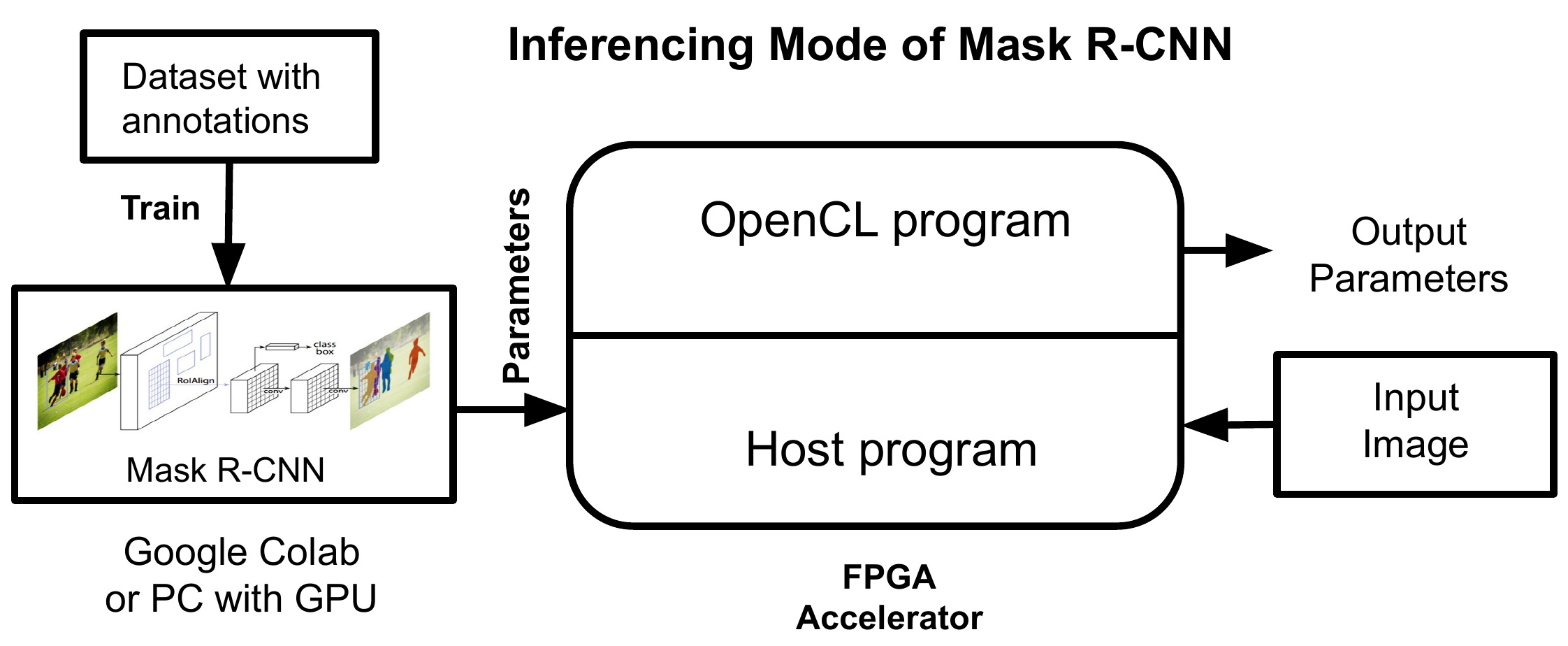}\]
  \caption{This figure shows how a Mask R-CNN acceleration could be done for an FPGA based deployment. The Mask R-CNN will get trained on a GPU/CPU based device and will then be converted into OpenCL equivalent code which can execute instructions parallely. The host code will handle the basic input and output of an image.}
  \label{fig:fpga}
\end{figure}

The practical implementation of this new scheme and its interfaces with the processing pipeline needs to be further explored. For this purpose we have investigated the use of FPGA platform and OpenCL languages. We have carried out a study for hardware acceleration of Mask R-CNN for a future implementation. Since the pulsar search pipeline runs on an accelerator,we have studied the possibility of using an FPGA platform. In our early investigation we found the OpenCL implementation of CNN available for DE-10 Nano FPGA board \footnote{\url{https://www.intel.com/content/www/us/en/developer/topic-technology/edge-5g/hardware/fpga-de10-nano}}. However more work needs to be done to map the different libraries and enhance it for Mask R-CNN. A proposed implementation of Mask R-CNN on hardware platform is shown in Figure \ref{fig:fpga}. We have also made a docker image of the basic Mask R-CNN software and will be enhancing it to deploy it over kubernetes clusters.

More investigation will have to be performed with a larger dataset. In the future study we will have more quantitative study with multiple classes.  We will also be investigating the image pixelization aspects in the future work. 

We would like to use real telescope data for our further investigations. In our future work, we will be investigating the Masking logic for identifying overlapping instances of butterfly pattern in an RFI background. Modern radio telescope data analysis faces challenges in the form of Radio Frequency Interference (RFI). 
The effect of RFI on our scheme requires future investigation. Normally RFI can be dealt with by capturing data in an RFI free observatory sites, removing the RFI prone datasets or by replacing the RFI affected set by benign random or median values \citep{Buch_2018,Buch_2022,2016JAI.....541018B,2019JAI.....840006B}. We have also proposed in our earlier work \citep{9232191} a novel method to identify observation slots that are likely to be free from RFI. It proposes a routine analysis of the radio telescope incoming data streams in order to identify RFI free observation slots using machine learning techniques. The methodology basically makes use of statistical analysis and detecting outliers in the data to build a comprehensive database. This database can be analyzed to view the RFI trends with time. The potential RFI free slots can be predicted using LSTM (Long Short Term Memory) techniques \citep{Misra2007CoSMOSPO, Misra2008SpreadSpectrum}.

The current work was taken up as a research activity towards identifying new algorithms and techniques for future implementations.

\section{Conclusion} 
\label{sec:conclusion}


The paper showed the potential of a machine learning algorithm Mask R-CNN for detecting candidate images in a pulsar search pipeline. We have tested this concept with a set of simulated data and shown the results here. We have also investigated the aspects of annotation for the training and presented a brief discussion. We have presented these details from the perspective of future improvements and hardware implementation. Our training made use of the cloud based computational infrastructure provided by Google. The entire work details, codes developed, data images used, output products including the neural network weight files are documented and available via github\footnote{\url{https://github.com/tprabu2000/Shashank-S-Bhat-Investigation-of-a-Machine-learning-methodology-for-the-SKA-pulsar-search-pipeline}}.  We anticipate the future pulsar search and similar candidate detection will benefit from this work. 

\section*{Acknowledgements}
The authors would like to thank the anonymous referee whose comments have vastly enhanced the quality of this manuscript. The authors would also like to thank Raman Research Institute, University of Manchester, PES University and BITS Goa for their support.

\setlength{\tabcolsep}{20pt}
\setlength{\columnsep}{2cm}
\renewcommand{\arraystretch}{1.5}

\bibliography{bibliography}
\end{document}